\documentclass[aps,prl,twocolumn,superscriptaddress]{revtex4}
\usepackage{graphicx,epsf}
\usepackage{amsmath}

\newcommand{\ybco}{YBa$_2$Cu$_3$O$_{6+\delta}$}
\newcommand{\ybcoa}{YBa$_2$Cu$_3$O$_{6.6}$}
\newcommand{\ybcob}{YBa$_2$Cu$_3$O$_{6.85}$}
\newcommand{\lsco}{La$_{2-x}$Sr$_x$CuO$_4$}
\newcommand{\bscco}{Bi$_2$Sr$_2$CaCu$_2$O$_{8+\delta}$}
\newcommand{\lbco}{La$_{2-x}$Ba$_x$CuO$_4$}
\newcommand{\lbcoo}{La$_{15/8}$Ba$_{1/8}$CuO$_4$}
\newcommand{\ccoc}{Ca$_{2-x}$Na$_x$CuO$_2$Cl$_2$}

\begin{document}

\title{
Spin excitations in fluctuating stripe phases of doped cuprate superconductors
}

\author{Matthias Vojta}
\affiliation{\mbox{Institut f\"ur Theorie der Kondensierten Materie,
Universit\"at Karlsruhe, 76128 Karlsruhe, Germany}}
\author{Thomas Vojta}
\affiliation{Physics Department, University of Missouri - Rolla, Rolla, MO 65409, USA}
\author{Ribhu K. Kaul}
\affiliation{Physics Department, Duke University, Science Drive, Durham, NC 27708, USA}
\affiliation{\mbox{Institut f\"ur Theorie der Kondensierten Materie,
Universit\"at Karlsruhe, 76128 Karlsruhe, Germany}}
\date{April 5, 2006}

\begin{abstract}
Using a phenomenological lattice model of coupled spin and charge modes,
we determine the spin susceptibility in the
presence of fluctuating stripe charge order.
We assume the charge fluctuations to be slow compared to those of the spins,
and combine Monte Carlo simulations for the
charge order parameter with exact diagonalization of the spin sector.
Our calculations unify the spin dynamics of both static and fluctuating stripe phases
and support the notion of a universal spin excitation spectrum
in doped cuprate superconductors.
\end{abstract}

\maketitle

%%%%%%%%%%%%%%%%%%%%%%%%%%%%%%%%%%%%%%%%%%%%%%%%%%%%%%%%%%%%%%%%%%%%%%%

A key challenge in the field of high-$T_c$ superconductivity is to
separate universal from non-universal properties.
For spin fluctuations, believed to be the glue that binds the Cooper pairs,
this issue is controversial:
early neutron scattering experiments had established the existence of
a ``resonance peak'', corresponding to a spin collective mode at the
antiferromagnetic wavevector, for certain cuprate families
\cite{respeak1,respeak2,respeak3},
while in others stripe-like spin and charge modulations were detected
\cite{lsco,waki,mook,pnas}.
(Signatures of charge order, likely pinned
by impurities, have been observed also in scanning tunneling microscopy (STM)
experiments on \bscco\ \cite{ali} and \ccoc\ \cite{hana}.)
Recent experiments have mapped out the spin excitations in various cuprates
over a large range of energies \cite{jt04,hinkov,buyers,hayden,face},
with remarkable results:
(i) In stripe-ordered \lbcoo\ (LBCO)
two excitation branches have been found, with
the high-energy branch above a ``resonance'' well described
by the spectrum of a gapped spin ladder \cite{jt04}.
(ii) \ybco\ shows incommensurate
excitations below the resonance energy \cite{hinkov,buyers,hayden}.
These results point toward a low-temperature spin excitation spectrum being
{\em universal} among the cuprate families at intermediate energies \cite{univ},
namely an ``hour-glass'' spectrum with a high-intensity peak at wavevector $(\pi,\pi)$ and
both downward and upward dispersing branches of excitations.

However, a unified theoretical description for the spin dynamics is lacking.
The spin excitations in LBCO \cite{jt04} appear to be well explained
within a model of static stripes \cite{MVTU,GSU,lorenzana}, where weak magnetic order
exists on top of a bond-ordered (i.e. dimerized) background.
In contrast, neutron scattering in \ybco\ was modelled using RPA-type
calculations \cite{rpa1,rpa2} which, however, rely on details of the band structure
and are not able to describe ordered states.
A key question is whether the neutron scattering data on \ybco\ can be understood
in a stripe picture as well -- as no static order has been detected, stripes have
to be fluctuating in space and time here.
Controversial experimental viewpoints on this have been put forward \cite{mook,hinkov,univ}.

The purpose of this letter is to show that {\em fluctuating stripes}
\cite{ssrmp,stevek,hassel} lead to spin excitations very similar to those observed
in the experiments,
thus providing a unified account of the collective mode dynamics in
the cuprates.
While large, weakly fluctuating, stripe domains are consistent with
the results in \lsco\ and \lbco,
a mixture of stripe and checkerboard structures
is required to explain the experimental data on \ybcob\ \cite{hinkov}.

%%%%%%%%%%%%%%%%%%%%%%%%%%%%%%%%%%%%%%%%%%%%%%%%%%%%%%%%%%%%%%%%%%%%%%%

{\it Lattice order parameter theory.}
We employ a phenomenological model of coupled spin and charge
fluctuations \cite{SNS} where the spin incommensurabilities are driven by
inhomogenieties in the charge sector;
this is supported, e.g., by experiments on stripe-ordered \lsco, where the
charge order sets in at a higher temperature than the spin order.
On a microscopic scale, the influence of the charge order on the spin sector
can be understood as spatial modulations of both spin densities and
magnetic couplings \cite{MVTU,GSU}.
(Additional collective degrees of freedom with zero wavevector,
e.g., pairing fluctuations, will not qualitatively modify our results.)
The goal of our work is to describe well-defined collective modes,
%in strongly correlated materials,
hence we neglect the continuum of single-particle excitations
and the associated collective mode damping.

The action of our Landau theory has the form
$\mathcal{S} = \mathcal{S}_\varphi + \mathcal{S}_\psi + \mathcal{S}_{\varphi\psi}$,
where $\mathcal{S}_\varphi (\mathcal{S}_\psi$) describe the spin (charge) fluctuations,
and $\mathcal{S}_{\varphi\psi}$ couples the two.
We assume a dominant antiferromagnetic interaction,
and so employ a lattice $\varphi^4$ theory for the spin fluctuations
at the {\em commensurate} wavevector ${\vec Q} = (\pi, \pi)$,
\begin{eqnarray}
\mathcal{S}_\varphi =
\int \! d \tau \sum_j \left[ (
\partial_\tau \vec\varphi_j)^2 +
s \vec\varphi_j^2 \right]
+  \sum_{\langle j j' \rangle} \! c^2
(\vec\varphi_j \!-\! \vec\varphi_{j'})^2 + \mathcal{S}_4
\nonumber
\end{eqnarray}
with $\mathcal{S}_4$ being the quartic self-interaction term.
%(The coupling to the charge stripes can then shift the minimum energy
%of the spin excitations to the incommensurate wavevector dictated by
%the charge order.)
The real order parameter $\vec\varphi_j$ and the spins $\vec S_j$ on
the sites $j$ of the square lattice are related through
$
\vec S_j \propto e^{i {\vec Q} \cdot {\vec r}_j} \vec\varphi_j
$.

Turning to the charge sector, we note that
microscopic calculations have indicated a tendency towards states with stripe-like
charge ordering \cite{doug,vs}, but
states with two-dimensional (2d) ``checkerboard'' modulations
closely compete in energy \cite{mv02,plaq2,seib00}.
We employ {\em two} complex order parameter fields $\psi_{x,y} ({\vec r}, \tau)$
which measure the amplitude of horizontal and vertical stripe order
at wavevectors ${\vec K}_{x,y}$.
Checkerboard order then implies {\em both} $\psi_x$ and $\psi_y$ non-zero.
In a situation with fluctuating charge order, the balance between stripes and
checkerboard (which depends on microscopic details \cite{mv02}) is controlled by
a repulsion or attraction between $\psi_x$ and $\psi_y$.
%i.e., the two order parameters are clearly not independent.
The complex phase of $\psi_{x,y}$ represents the sliding degree of freedom
of the density wave and distinguishes between bond- and site-centered stripes.
In our simulations we concentrate on ${\vec K}_{x}\!=\!(\pi/2,0)$ and
${\vec K}_y\!=\!(0,\pi/2)$, i.e., a charge modulation period of 4 lattice
spacings; modulations at these wavevectors have been observed both in
neutron scattering \cite{lsco,waki,mook,pnas} and STM \cite{ali,hana},
in particular near doping 1/8 where stripe order is most robust.
The real field
$
Q_x ({\vec r} ) = {\rm Re} \psi_x ({\vec r}) e^{i {\vec K}_x \cdot {\vec r}}
$
(similarly for $Q_y$) measures the modulation of both the charge density
(for $\vec r$ on sites) and bond order (i.e., kinetic energy or pairing amplitude,
for $\vec r$ on bonds).
We choose signs such that
$\delta\rho({\vec r}_j) = Q_x + Q_y$ is the deviation of the local {\em hole} density
from its spatial average.
The all-important couplings between spin and charge fluctuations have to be
of the form $\lambda Q \vec\varphi^{\,2}$ due to the underlying SU(2) symmetry.
Guided by the lattice models \cite{MVTU,GSU} we choose \cite{SNS}
\begin{eqnarray}
\mathcal{S}_{\varphi\psi} &=& \int d \tau \sum_j \Bigl[
\lambda_1 Q_x ({\vec r}_j ) \vec\varphi_{j}^2 + \lambda_2 Q_x
({\vec r}_{j+x/2})
\vec\varphi_{j} \vec\varphi_{j+x} \nonumber\\
&+& \lambda_3 Q_x ({\vec r}_j)
\vec\varphi_{j-x} \vec\varphi_{j+x} + \lambda_4 Q_x ({\vec
r}_{j+y/2}) \vec\varphi_{j} \vec\varphi_{j+y} \Bigr] \nonumber \\
&+&  \Bigl[ x \leftrightarrow y \Bigr].
\nonumber
\end{eqnarray}
$\lambda_1>0$ implements the correlation between the
on-site charge density and the amplitude of the spin fluctuations,
while $\lambda_{2-4}$ ensure that the effective first- and
second-neighbor exchange constants modulate along with the bond order;
the antiphase domain wall properties of the stripes \cite{pnas,doug}
are reflected in the positive sign of $\lambda_{2,3}$.

For constant $\psi_{x,y}$ the action
$\mathcal{S}_\varphi + \mathcal{S}_{\varphi\psi}$
is a theory for magnetic modes in a background of static
charge order.
For sufficiently large $\lambda$ couplings, the minimum energy of the
$\varphi$ fluctuations will be shifted away from $(\pi,\pi)$
to the incommensurate wavevector dictated by the charge order
(this is a non-perturbative effect!),
with the spin order remaining collinear.
Results \cite{SNS} for the spin fluctuation spectrum in the presence of
static stripes are in excellent agreement
with results on LBCO \cite{jt04}.

%%%%%%%%%%%%%%%%%%%%%%%%%%%%%%%%%%%%%%%%%%%%%%%%%%%%%%%%%%%%%%%%%%%%%%%

\begin{figure}[!t]
\epsfxsize=2.9in
\epsffile{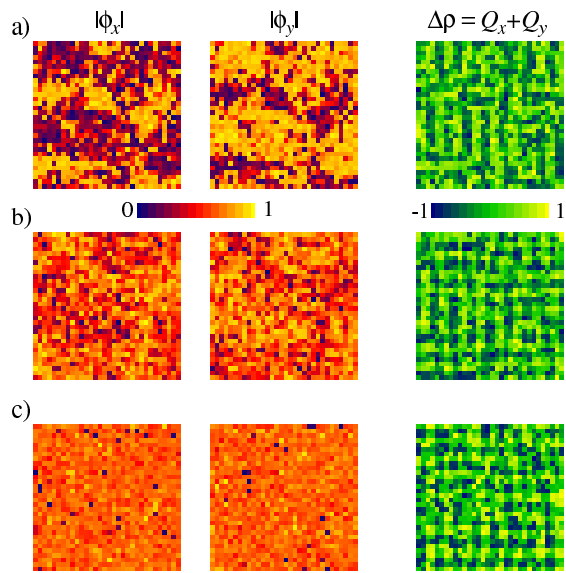}
\caption{(color online)
Snapshots of the charge order parameters $|\psi_{x,y}|$ and
the resulting charge modulation $(Q_x + Q_y)$, obtained from
MC simulations for bond-centered fluctuating stripes on $32^2$ sites,
using a $\psi^6$ action ${\cal S}_\psi$, see text.
a) Strong repulsion between $\psi_x$ and $\psi_y$, sharp domain walls.
b) Weak repulsion, smooth domain walls with checkerboard structure.
c) Weak attraction, fluctuating checkerboard order.
\vspace*{-8pt}
}
\label{fig:cfg}
\end{figure}

%%%%%%%%%%%%%%%%%%%%%%%%%%%%%%%%%%%%%%%%%%%%%%%%%%%%%%%%%%%%%%%%%%%%%%%

{\it Fluctuating charge order.}
We now turn to ${\cal S}_{\psi}$:
For slowly fluctuating charge order it is useful to think about
snapshots of the charge configuration.
Defining an O(4) field $\psi=(\psi_x,\psi_y)$
we can discuss physically distinct spatial fluctuations:
(i) Fluctuations of the complex phases of $\psi_{x,y}$ are stripe
or checkerboard dislocations.
(ii) Fluctuations between areas of dominant $\psi_x$ or $\psi_y$
represent domain walls between horizontal and vertical stripes.
(iii) Variations in $|\psi|$ are amplitude fluctuations in the
local charge order.
Depending on the particular form of ${\cal S}_{\psi}$, these fluctuations
will have different importance.
Regarding amplitude fluctuations two extreme cases come to mind:
(a) a standard $\psi^4$ theory with a ``soft'' order parameter, which
has rather large amplitude fluctuations, and
(b) a ``hard'' order parameter theory with a fixed-length constraint,
$|\psi|^2={\rm const}$.
Although microscopically amplitude fluctuations are present,
existing approximate results for Hubbard or $t$-$J$ models are inconclusive
with regard to their importance.
Experimentally, STM results \cite{hana} on \ccoc\ indicate a spatially
disordered arrangement of stripe segments and more 2d ``tiles'',
with the amplitude of these local modulations fluctuating rather little.

In our simulations, we have employed various forms for the charge action $\mathcal{S}_\psi$,
with different amounts of amplitude fluctuations. Most useful is
a $\psi^4$-type theory for the O(4) field $\psi$, supplemented
by a positive $\psi^6$ term:
\begin{eqnarray}
&&\mathcal{S}_{\psi} = \int d \tau d^2 {\bf r} \Bigl[
\left| \partial_\tau \psi_x \right|^2 +
\left| \partial_\tau \psi_y \right|^2 +
s_x |\psi_x|^2 + s_y |\psi_y|^2 \nonumber\\
&&+
 c_{1x}^2 \left| \partial_x \psi_x \right|^2 + c_{2x}^2 \left|
\partial_y \psi_x \right|^2 + c_{1y}^2 \left|
\partial_y \psi_y \right|^2 + c_{2y}^2 \left|
\partial_x \psi_y \right|^2
\nonumber \\
&&+
u_1 \psi^4 + u_2 \psi^6
+ v |\psi_x|^2 |\psi_y|^2
+ w \left( \psi_x^4 \!+\! \psi_x^{\ast 4}
\!+\! \psi_y^4 \!+\! \psi_y^{\ast 4} \right) \Bigr]
\nonumber
%\label{psiact}
\end{eqnarray}
with $\psi^2\!\equiv\!|\psi_x|^2\!+\!|\psi_y|^2$.
A combination of $u_1\!<\!0$ and $u_2\!>\!0$ suppresses amplitude fluctuations
of $\psi$. For $c_{1x}\!=\!c_{1y}$, $c_{2x}\!=\!c_{2y}$, $s_x\!=\!s_y$, and $v\!=\!w\!=\!0$,
the action has O(4) symmetry.
The $w$ term selects between bond-centered and site-centered stripes.
The important quartic $v |\psi_x|^2 |\psi_y|^2$ term regulates
the repulsion or attraction between horizontal and vertical stripes,
i.e., it determines whether the character of the order will be
one-dimensional (stripe, for $v>0$) or two-dimensional (checkerboard, for $v<0$).

To simplify the treatment of $\mathcal{S}_\varphi + \mathcal{S}_{\psi} + \mathcal{S}_{\varphi\psi}$,
we assume that fluctuations in the charge sector are slow compared to
those in the spin sector, and we neglect the feedback of the spins
on the charges.
This leads to an adiabatic (Born-Oppenheimer) approximation for the
coupled dynamics, and allows to treat the charge fluctuations by {\em classical}
lattice Monte Carlo (MC) simulations.
For each configuration of the $\psi_{x,y}$, the remaining theory
$\mathcal{S}_\varphi + \mathcal{S}_{\varphi\psi}$ (at the Gaussian level, ${\cal S}_4=0$)
is quadratic in the $\varphi$ fields and can be diagonalized on lattices up
to $64^2$ sites.
(Neglecting ${\cal S}_4$ is justified in spin-disordered phases.)
We employ a standard Metropolis algorithm with single-site updates at a
finite effective temperature ($T=1$) to simulate ${\cal S}_{\psi}$ in a regime
where the correlation length $\xi$ is between 5 and 50 lattice spacings \cite{param}.
The spin susceptibility $\chi''(\vec q,\omega)$ is obtained by averaging its value
over typically 20 MC charge configurations, with $10^5$ -- $10^6$ MC steps between
two measurements.

%%%%%%%%%%%%%%%%%%%%%%%%%%%%%%%%%%%%%%%%%%%%%%%%%%%%%%%%%%%%%%%%%%%%%%%

\begin{figure}[!t]
\epsfxsize=3.2in
\epsffile{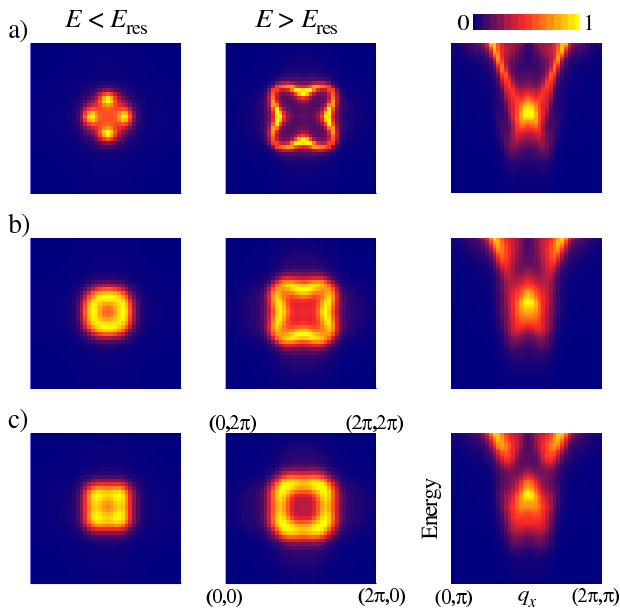}
\caption{(color online)
Dynamic susceptibility $\chi''({\vec q},\omega)$
for bond-centered fluctuating stripes \cite{site} on $40^2$ sites.
%(Results for site-centered stripes are similar.)
Left/Middle: cuts at a constant energy, slightly below/above the
resonance energy, $E_{\rm res}$, as function of momentum.
Right: cuts along $(q_x,\pi)$ as function of $q_x$ and energy,
showing the universal ``hour-glass'' spectrum.
The couplings are $\lambda_1 = \lambda_3/2 = 5 E_{\rm res}/|\psi|_{\rm typ}$,
$\lambda_2=\lambda_4=0$.
a) Strong repulsion between $\psi_x$ and $\psi_y$, correlation length $\xi\approx 30$.
b) Weak repulsion,  $\xi\approx 20$.
c) Weak attraction, $\xi\approx 20$.
\vspace*{-8pt}
}
\label{fig:chicol1}
\end{figure}

%%%%%%%%%%%%%%%%%%%%%%%%%%%%%%%%%%%%%%%%%%%%%%%%%%%%%%%%%%%%%%%%%%%%%%%

{\it Numerical results.}
Typical snapshots of the two charge order parameters $\psi_{x,y}$ and the
resulting charge configuration, for different values of the stripe interaction $v$,
are shown in Fig.~\ref{fig:cfg}.
Let us now discuss our results for the dynamic spin susceptibility,
$\chi''(\vec q,\omega)$, as measured in inelastic neutron scattering.
Starting from ordered stripes \cite{SNS},
we found that amplitude fluctuations of $\psi$ rather quickly destroy the
incommensurate spin response; for a standard $\psi^4$ theory this happens already
within the ordered phase.
In contrast, the spin sector turns out to be less sensitive to phase fluctuations
of the stripe order.
Interpreted microscopically, this means that incommensurate spin response requires
well-formed stripe segments with a length of order 10 lattice spacings.
We have therefore focussed on versions of ${\cal S}_\psi$ with small
amplitude fluctuations, and carried out large-scale simulations for
various couplings and correlation lengths of the $\psi_{x,y}$.

\begin{figure}[!b]
\epsfxsize=3.2in
\epsffile{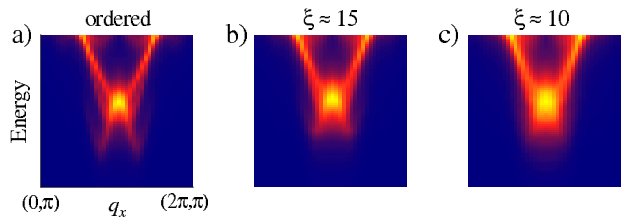}
\caption{(color online)
Evolution of $\chi''({\vec q},\omega)$
with decreasing stripe correlation length $\xi$, for a situation
with strong repulsion between $\psi_x$ and $\psi_y$,
as in Fig.~\protect\ref{fig:chicol1}a.
%The incommensurate spin response in the lower branch is most pronounced for large $\xi$,
%resulting in an ``hour-glass'' or ``X''-shaped spectrum.
%In contrast, small $\xi$ smears the resonance peak and yields a more ``Y''-shaped spectrum
%with little momentum dependence over a significant energy range around $E_{\rm res}$.
\vspace*{-8pt}
}
\label{fig:chicol4}
\end{figure}

Results for the dynamic susceptibility,
corresponding to the situations in Fig.~\ref{fig:cfg},
are shown in Fig.~\ref{fig:chicol1}.
The right panels show that a common feature of all spectra is an ``hour-glass''
(or ``X-shaped'') spectrum with a strong ``resonance'' peak at $(\pi,\pi)$
and energy $E_{\rm res}$.
The downward dispersing lower branch is most pronounced for well-defined stripes, i.e.,
large $\xi$; it is progressively smeared out with decreasing $\xi$ (Fig.~\ref{fig:chicol4}).
Very recent neutron scattering results \cite{hinkov2}
indicate a ``Y-shaped'' response in the pseudogap state above $T_c$ in
underdoped \ybcoa, with little dispersion at low energies --
these data seem to be consistent with fluctuating short-range stripe segments,
Fig.~\ref{fig:chicol4}c.

We have studied in detail the crossover from a strictly ``stripy''
situation with a strong repulsion between $\psi_x$ and $\psi_y$
(Figs.~\ref{fig:cfg}a, \ref{fig:chicol1}a, \ref{fig:chicol4}) to
a 2d checkerboard regime with attraction between
$\psi_x$ and $\psi_y$ (Figs.~\ref{fig:cfg}c, \ref{fig:chicol1}c).
Significant differences occur in the lower branch
(left panels of Fig.~\ref{fig:chicol1}).
For large domains of horizontal or vertical stripe order, Fig.~\ref{fig:chicol1}a,
well-defined peaks occur along $(q_x,\pi)$ and $(\pi,q_x)$,
as observed in LBCO.
Interestingly, with increasing volume fraction of checkerboard domain walls,
Fig.~\ref{fig:chicol1}b,
the neutron response is both smeared and enhanced along the $q$-space diagonals,
resulting in a quasi-2d dispersion of the downward branch --
this is strikingly similar to data on \ybcob\ \cite{hayden}.
Finally, in a checkerboard regime the low-energy spin excitations occur
along the diagonals (Fig.~\ref{fig:chicol1}c).
Based on these results and STM data \cite{hana} we predict that such a
neutron response should be observable in \ccoc.

The $q$-space structure of the upward dispersing branch
(middle panels of Fig.~\ref{fig:chicol1})
changes less from the stripe to the checkerboard regime;
it is strongly anisotropic only in the stripy situation of Fig.~\ref{fig:chicol1}a
where it resembles the spectrum of two-leg ladders.
Focussing on the right panels of Fig.~\ref{fig:chicol1},
we further observe that larger checkerboard regions tend to suppress
the upper branch right above the resonance peak,
it re-appears only at somewhat higher energies.
This is in remarkable agreement with neutron data on \ybcob\ \cite{face},
and can be easily understood:
For perfect checkerboard order the low and high-energy response are
separated by a large gap \cite{SNS},
and our simulations interpolate between stripes and checkerboard.
(An alternative interpretation of the data of Ref.~\onlinecite{face}
within RPA is in Ref.~\cite{rpa2}.)

\begin{figure}[!t]
\epsfxsize=3.2in
\epsffile{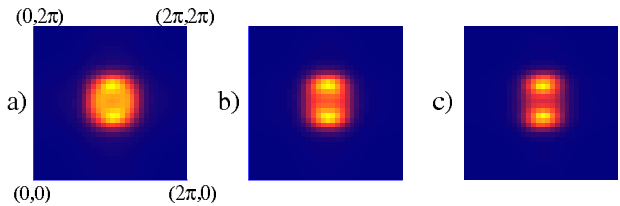}
\caption{(color online)
Dynamic susceptibility slightly below the resonance energy,
for fluctuating stripes in the presence of an in-plane anisotropy.
From a) to c) the anisotropy is increasing:
the ratio of the gradient coefficients for $\psi_{x,y}$ in the charge action
$\mathcal{S}_\psi$ is a) 1.005, b) 1.01, c) 1.02.
\vspace*{-8pt}
}
\label{fig:chian1}
\end{figure}

To model detwinned \ybcob\ \cite{hinkov},
it is necessary to include an in-plane anisotropy to account
for the orthorhombic distortions.
Assuming the anisotropy to be small, it will mainly influence the
low-energy charge fluctuations.
We have therefore carried out simulations where the
$\psi_x$ and $\psi_y$ order parameters had different mass and/or different
gradient terms (i.e. velocities).
Sample results for the downward dispersing branch are shown in
Fig.~\ref{fig:chian1}, which are in reasonable agreement with the
data of Hinkov {\em et al.} \cite{hinkov}.
The anisotropy decreases at higher energies (not shown).

Let us note that our adiabatic approximation to treat
$\mathcal{S}_\varphi + \mathcal{S}_{\psi} + \mathcal{S}_{\varphi\psi}$
cannot distinguish between slowly fluctuating and time-independent disordered
stripes (e.g., a ``stripe glass'' pinned by impurities).
Physically, these two situations will
indeed yield very similar spin excitations;
a clear-cut distinction will require a direct probe of the charge modes.

%%%%%%%%%%%%%%%%%%%%%%%%%%%%%%%%%%%%%%%%%%%%%%%%%%%%%%%%%%%%%%%%%%%%%%%

{\it Conclusions.}
We have determined the spin excitations in the presence of fluctuating
stripe charge order.
We obtain an incommensurate spin response provided that
(i) charge order fluctuates predominantly in phase rather than
in amplitude, and
(ii) the charge correlation length is at least 10 lattice spacings.
(Assuming a collective mode velocity of 50 meV this roughly translates
into THz fluctuation frequencies.)
In addition, we found that an increasing volume fraction of stripe domain
walls with checkerboard structure leads to quasi-2d spin fluctuations
as observed in \ybcob.

Our calculations thus support the notion of a universal
spin excitation spectrum at intermediate energies in the cuprates,
arising from stripe-like charge-density fluctuations.
This brings us closer to a unified description of the collective
excitations in the high-$T_c$ materials.

%%%%%%%%%%%%%%%%%%%%%%%%%%%%%%%%%%%%%%%%%%%%%%%%%%%%%%%%%%%%%%%%%%%%%%%

% \acknowledgments

We thank A. V. Balatsky, W. Byers, B. Keimer, V. Hinkov, C. Pfleiderer, A. Rosch,
J. Tranquada, and G. Uhrig for discussions,
and especially S. Sachdev for collaborations on related work.
This research was supported by
%the DFG Center for Functional Nano\-struc\-tures and
the Virtual Quantum Phase Transitions Institute (Karlsruhe),
by NSF Grants PHY99-0794 (KITP Santa Barbara), DMR-0339147 (TV), DMR-0103003 (RKK),
the DAAD (RKK), and the Research Corporation (TV).

%%%%%%%%%%%%%%%%%%%%%%%%%%%%%%%%%%%%%%%%%%%%%%%%%%%%%%%%%%%%%%%%%%%%%%%


\begin{thebibliography}{}
\bibitem{param}
$\mathcal{S}_\psi$ parameters were $u_1\!=\!-1\ldots-2$, $u_2\!=\!0.1$,
$c_{1,2;x,y}\!=\!0.15\ldots0.25$, $s_x\!=\!s_y\!=\!-6\ldots-1$,
and $w\!=\!0.05$ resulting in bond-centered stripes \cite{site}.
The parameter $v$, selecting between stripes and checkerboard,
was $v=0.2, 0.04, -0.02$ for Figs. \ref{fig:cfg}a,b,c, respectively.
The velocity in ${\cal S}_{\varphi}$ is $c=100$ meV, resulting
in $E_{\rm res} \approx 40$ meV.
Fixing the typical order parameter amplitude $|\psi|_{\rm typ}$,
$s$ and the $\lambda$ are chosen such that a perfectly
charge-ordered reference state has gapless incommensurate spin excitations as in
Ref.~\onlinecite{SNS}.
\bibitem{site}
For site-centered stripes, fluctuations are more effective
in destroying the incommensurate spin response.
\end{thebibliography}
\end{document}